\documentclass[english,aps,prstper,reprint,showpacs,titlepage]{revtex4-2} 

\usepackage[T1]{fontenc}	
\usepackage{geometry}		
\usepackage{orcidlink}
\usepackage[noorphans,vskip=0.5ex]{quoting}
\usepackage{graphicx}
\usepackage[above,below]{placeins}	
\usepackage{times}
\usepackage{booktabs}
\usepackage{hyperref}  
\hypersetup{colorlinks=true,urlcolor=blue,citecolor=blue,linkcolor=blue}   
\begin{document}

\begin{titlepage}
  \title{"I don't have a good metric for what other people think:" Unpacking the double empathy problem in neurodivergent physics problem-solving}

    \author{Liam~G.~McDermott$^1$$^,$$^2$ \orcidlink{0000-0002-0594-0486}}
    \author{Frank T. Dachille$^3$ \orcidlink{0000-0002-0219-6284}}
    \author{George R. Keefe$^4$}
    \author{Mason~D.~Moenter$^5$ \orcidlink{0009-0004-0177-2210}}
    \author{Daryl McPadden$^3$ \orcidlink{0000-0003-3739-4448}}
    \author{Erin M. Scanlon$^1$ \orcidlink{0000-0001-8672-7187}}

    \affiliation{$^1$Department of Physics, University of Connecticut $-$ Avery Point, 1084 Shennecossett Rd, Groton, CT, 06340}
    \affiliation{$^2$ Department of Physics, Durham University, Durham, UK}
    \affiliation{$^3$Department of Physics \& Astronomy, Michigan State University, 567 Wilson Rd, East Lansing, MI 48824}
    \affiliation{$^4$Physics and Astronomy Department, Rutgers University, 136 Frelinghuysen Rd, Piscataway, NJ 08854}
    \affiliation{$^5$Department of Physics \& Astronomy, Texas A\&M University, 578 University Drive College Station, TX 77843}

\begin{abstract}

\end{abstract}
  

\begin{abstract}

Learning is a social act and requires communication between interlocutors. Learning also is a process that takes place in a socially constructed neuro-normative system. These neuro-normative systems can marginalize neurodivergent (i.e., autistic, dyslexic, epileptic) students, who oftentimes have non-normative socio-cognitive differences. In this paper, we investigate 3 neurodivergent physics students' experiences solving physics problems and in navigating physics culture. We use a Double-Empathy Problem lens to uncover how communication, particularly communication breakdowns, can manifest for neurodivergent students in physics. We find that the 3 neurodivergent students, experience communication breakdowns in their interactions with peers and instructors, and in interpreting problem statements. Furthermore, we find that participants experience fear of being misinterpreted or feeling as though they are viewed as stupid on account of communication breakdowns. We advocate for the need for a more neuro-inclusive physics culture and suggest actions that can be done by instructors to create more inclusive spaces in physics.  

\clearpage

\end{abstract}
  \maketitle
  
\end{titlepage}

\begin{center}
    \textbf{I. INTRODUCTION}
\end{center}

Society is made of many intersecting and interdependent neuro-normative systems (systems that privilege neurotypical ways of cognition, behavior, and socialization), \cite{Yergeau2018,Radulski2021,Hennekam2024,Field2025},  including higher education generally \cite{Mellifont2023,Kennedy2025}. Neurodivergent physics students report they do physics differently than their neurotypical peers \cite{McDermott2025}, which, when inside a system that marginalizes non-normative ways of doing physics, can have negative impacts on student recognition and belonging in physics \cite{McDermott2024b}. Researchers have investigated the social dimension of neurodivergent STEM student experiences in group-work \cite{Salvatore2024,Andreassen2017,Pfieffer2023}, presentations \cite{Hand2023}, active learning environments \cite{James2020}, and interactions with peers \cite{Phillips2024}. Findings across these studies indicate that there is indeed a social aspect of higher education STEM classes that marginalizes neurodivergent students.

Despite the number of studies that indicate a marginalizing effect caused by social expectations in neuro-normative higher education settings, very few use a neurodiversity-specific framing to investigate the phenomenon. We use the Double Empathy Problem (DEP) as such a framework to focus on neurodivergent-neurotypical socialization \cite{Milton2012, Milton2022}, that is, the communication between neurodivergent and neurotypical people within a neuro-normative system. As a framework, the DEP both identifies moments of communication breakdowns between neurodivergent \& neurotypical (cross-neurotype) interlocutors and interprets these moments as a strengths-based, neurodiversity-affirming theory. Importantly, the DEP as a framework focuses on how cross-neurotype communication shifts between contexts and as systems of power change \cite{Kilgalon2026}. In this paper, we use the DEP to examine the experiences of individuals as an avenue for critiquing systems of power. As a part of a larger study into neurodivergent student physics problem-solving, we use the DEP to frame our analysis of a series of think-aloud interviews with 3 neurodivergent physics students, and to help answer the following research questions:
\begin{enumerate}
    \item \textit{What are the ways neurodivergent students notice communication breakdowns in physics problem-solving contexts?} 
    \item \textit{Where do communication breakdowns occur for neurodivergent students solving physics problems?}
\end{enumerate}
In addition, we add to the literature by providing a systems-level and physics problem-solving context to the DEP; examining how neurodivergent students navigate social aspects of problem-solving in a neuro-normative system.

\begin{center}
    \textbf{II. CONCEPTUAL FRAMEWORKS}
\end{center}

We use two frameworks to guide the design, analysis, and dissemination of this study. We use the neurodiversity approaches \cite{Dwyer2022} framework to guide the design and dissemination of this study; and use the DEP to inform the analysis, specifically how we conceptualize and identify DEP events in the data. 
\newpage
\begin{center}
    \textbf{A. Neurodiversity Approaches}
\end{center}

The neurodiversity approaches \cite{Dwyer2022} are a set of guidelines for the design, implementation, and dissemination of research that involves neurodivergent people. In the neurodiversity approaches framework, disability/neurodivergence arises from interactions between a person's innate characteristics and their environment/other people \cite{Bolte2021, Dwyer2022, McDermott2024a}. Furthermore, when designing research using the neurodiversity approaches, researchers should:
\begin{enumerate}
    \item Examine both individual- and systems-level effects of neurodivergence,
    \item Include neurodivergent people at all stages of research design, and
    \item Work to rectify the harm that research and researchers have done to neurodivergent people.
\end{enumerate}

In this study, we examine how, in a neuro-normative system, neurodivergent students' individual strengths may be masked, and how neuro-normative systems serve to marginalize students on account of socio-cognitive differences. We specifically use frameworks developed by neurodivergent people: the neurodiversity approaches and the DEP. This study is designed by neurodivergent people, and we privilege the voices of neurodivergent people through our participant interviews and use of member-checking \cite{McKim2023}. We hope to rectify some of the harms caused by research and other academic neuro-normative systems by including recommendations for further research and neuro-inclusive pedagogy. 

\begin{center}
    \textbf{B. The Double-Empathy Problem}
\end{center}

The DEP is a strengths-based theory that describes breakdowns in cross-neurotype communication resulting from socio-cognitive, experiential, and expectational differences \cite{Milton2012, Milton2022}. The DEP was created in direct response to deficit-based theories of autistic communication, such as Baron-Cohen's mind-blindness \cite{BaronCohen1990}. Ironically, the DEP does not formalize its use of empathy, opting against the constraints of cognitive and affective empathy \cite{Ekdahl2024, Uhlman2025}, meaning that we depart from previous physics education research (PER) conceptualizations of empathy \cite{Merrill2024, Hamdan2024}. Instead, the DEP uses a folk-definition of empathy, focusing on understanding (i.e., perception, relation, and connection) between two interlocutors \cite{Milton2021, Milton2022, Ekdahl2024}. This lack of formalization is purposeful, serving to make space for the heterogeneity of the ways people relate to and understand one another \cite{Ekdahl2024}. 

In the DEP, cross-neurotype communication breakdowns are the result of interactions between people "who hold different norms and expectations of each other" \cite{Milton2021} $(p.1)$. Breakdowns in communication, therefore, come not from inherent relational or empathic deficits in neurodivergent people, as in mind-blindness conceptualizations, but from people's neutral socio-cognitive differences, different life experiences, and expectations of communication, as communication is a two-way street.

When systems-level considerations are applied, we see that consistent breakdowns in communication have a damaging effect on neurodivergent participants, isolating neurodivergent people during childhood \cite{Mitchell2021}. Neuro-normative forms of communication are entrenched in higher education systems \cite{Tewes2024}. As such, neurodivergent people, operating under different modalities of communication, are systematically excluded, discriminated against, and stigmatized \cite{Tewes2024,Rubenstein2025}. We therefore use the DEP as a Critical framework (critiquing power and norms) to pay particular attention to systems-level effects of a neuro-normative culture on participants through a DEP lens.

We use DEP as a lens for interpreting communication breakdowns associated with differences in expectations, norms, and experiences. We specifically use an \textit{a priori} code we called a "DEP event." Using the DEP framework, we define a DEP event as: \textit{an instance where communication breaks down due to differences in expectations, norms, and experiences occurs, or where an action is taken because of the perception that a communication breakdown will occur.}

\begin{center}
    \textbf{III. METHODS AND POSITIONALITY}
\end{center}

\begin{center}
    \textbf{A. Data Collection and Demographics}
\end{center}

As a part of a larger study investigating neurodivergent student problem-solving in physics, we recruited participants using American Physical Society and university-specific listservs. We interviewed $N=17$ neurodivergent physicists in this study, and we report on $N=3$ of the participants' interviews. This was a deliberate choice by the authors to allow for a deeper dive into participants' narratives so that we could analyze communication from their standpoints as neurodivergent students. We chose the three participants (Ray, Moonbeam, and Milo), as they spent a great deal of time talking about communication in their interviews.

To collect the data, McDermott conducted a series of 2 think-aloud interviews and 1 semi-structured interview with the participants. The first think-aloud interview consisted of 5 introductory mechanics questions, while the second consisted of 3 introductory electricity and magnetism problems. The semi-structured interview consisted of 12 open-ended questions about experiences being a neurodivergent physics student. McDermott conducted all interviews using Webex.

We collected participant demographics by asking open-ended questions at the end of the first interview, in line with recommendations in previous work by the authors \cite{Moenter2025}. Participant demographics are given in Table \ref{tab:ParticipantDemographics}. 

\vspace{-3mm}

\begin{table}[h]
\caption{\label{tab:ParticipantDemographics}Participant demographics. All descriptions are taken from participants' own words.}
\begin{tabular}{l|l|l|c|l}
\toprule
\textbf{Pseudonym} & \textbf{Race} & \textbf{Gender} & \textbf{Year} & \textbf{Neurodivergence} \\
\midrule
Moonbeam   & \parbox[t]{1.8cm}{Latin, white}  & \parbox[t]{1.5cm}{Cis Woman}     & 3 & \parbox[t]{2.5cm}{ADHD, Autistic} \\ \hline
Ray        & \parbox[t]{1.8cm}{Puerto Rican, white} & \parbox[c]{1.5cm}{Trans Man}  & 4 & \parbox[t]{2.5cm}{ADHD, Autistic, OCD, Anxiety, Depression} \\ \hline
Milo       & \parbox[t]{1.8cm}{white}  & Trans Man  & 4 & \parbox[t]{2.5cm}{Autistic} \\ \hline
\bottomrule
\end{tabular}
\end{table}

\vspace{-3mm}
\newpage
\begin{center}
    \textbf{B. Analysis}
\end{center}

We used a thematic analysis method to analyze the transcripts \cite{Kushnir2025}. Authors 1-3 independently coded all transcripts. We used a deductive coding method, using the \textit{a priori} code "DEP event," outlined in section II.B. After coding the transcripts, Authors 1-3 met to discuss their codes and engaged in an interrater agreement process \cite{Cofie2022}. This process entailed 1 hour of debate between the authors regarding where in the data DEP events occurred and 3 hours of debate to define the patterns we report in section IV (what ways neurodivergent students notice communication breakdowns and where those breakdowns occur when solving physics problems). This process was led by McDermott, however given the power dynamics caused by him being the research lead, Dachille and Keefe predominantly led the discussion while McDermott asked guiding, clarifying, and analysis-honing questions. There were no instances where the 3 authors could not come to an agreement.


In addition to our analysis, we engaged in a member-checking process \cite{McKim2023}. Once the draft of this paper was finished, we sent the draft to all participants with the request to read through the paper and make comments regarding how they were represented in our analysis and how the findings lined up with their lived experience. We gave participants 2 weeks to respond with comments. 1 participant responded, and gave affirmative comments.

\begin{center}
    \textbf{C. Positionality}
\end{center}

The research team is neurodiverse, consisting of neurotypical, autistic, ADHD, OCD, and dyslexic individuals, strengthening our analysis as we have both insiders and outsiders to the neurodivergent community. The lead researcher, who conducted all interviews, shared neurotypes with multiple participants. The questions in our think-aloud interviews, however, are taken from textbooks, PhysPort \cite{PhysPort}, and neurotypical instructors. Having a neurodivergent interviewer allowed us to examine systemic effects caused by doing physics in a neuro-normative system, rather than due to miscommunication between a neurotypical interviewer and neurodivergent participants.

\begin{center}
    \textbf{IV. FINDINGS}
\end{center}

\begin{center}
    \textbf{A. What are the ways neurodivergent students notice communication breakdowns in physics problem-solving contexts?}
\end{center}

Throughout the three interviews, and across the three participants, we found multiple instances of communication breakdown, both directly during problem-solving, and through anecdotes shared during the semi-structured interview. For instance, while describing how he works with his peers, Milo states:
\begingroup
\addtolength\leftmargini{-0.16in}
\begin{quote}
    \textit{"[Two friends and I] had some trouble being- like, working together. Because if the three of us were in a class together, we would all be working together, and we would be going back and forth, just not understanding what the other person was saying."}
\end{quote}
\endgroup
Similarly, while solving a ranking problem, Ray double checks his answer against internet resources because he fears that the way he communicates his ideas are not the way that make sense to other people. He states: 
\begingroup
\addtolength\leftmargini{-0.16in}
\begin{quote}
\textit{"I'm checking my own brain against other people's brains ... I have a really hard time because a lot of things make sense to me. And then a lot of those things don't make sense to other people."}
\end{quote}
\endgroup
Both Ray and Milo find, through past experience, that the way they communicate about physics problem-solving is conflicting with the ways their peers or instructors communicate. Ray preempts his problem-solving by voicing communication concerns with other people even when there is a neurodivergent interviewer, no one has said that his answer doesn't make sense, he is not being assessed for correctness, and there are no other problem-solvers around. It seems that these communication breakdowns arise through neurodivergent differences in expectations and norms in what information is needed, the order in which information is presented, and the modality (i.e., mathematical reasoning, storytelling, analogy, drawing) of communication. This evidence of communication breakdown through differences in norms and expectations indicates that the DEP is present in neurodivergent physics problem-solving. We found that, while the DEP was present on an individual level (i.e., student-peer and student-instructor), there is a systemic effect related to the DEP present in physics (problem statement - student).

\begin{center}
    \textbf{B. Where do communication breakdowns occur for neurodivergent students solving physics problems?}
\end{center}

We found two spaces where the DEP is particularly prevalent in participants' physics problem-solving and in their experience in physics: interactions with peers and instructors, and interpreting problem-statements. 

\begin{center}
    \textit{1. Interactions with Peers and Instructors}
\end{center}

When the DEP arises between interlocutors in class, the communication breakdown that follows can be detrimental to students. Moonbeam describes her communication experience in class:
\begingroup
\addtolength\leftmargini{-0.16in}
\begin{quoting}
    "\textit{The professor is like, `You know what? I'm just gonna skip all this process. And does everyone understand how we get here?' And then you see all these people nodding their heads and I'm like, `But I don't. I don't get it.'"}
\end{quoting}
\endgroup

Moonbeam indicates that her professor has certain expectations of students, and because Moonbeam does not match those expectations, she is excluded from the communication of information in class.

Experiencing communication breakdown with peers and instructors has a negative impact on Moonbeam's view of herself as a competent physicist. Following her description of experiencing communication breakdowns, Moonbeam states, "\textit{I get very nervous all the time that again, people will think I'm stupid, especially my peers and my professors,}" a sentiment that is echoed across the 3 participants.

Talking about a conversation she had with her peers about coordinate systems, Moonbeam describes that her peers are able to describe their reasoning in problem-solving in a way that others understand and agree with. In contrast, Moonbeam states, "\textit{[I feel] like I can't even put parts of my thought process into words, while [my peers] can,}" indicating a communication breakdown between her and her peers. It is not that Moonbeam is an unpracticed or unskilled communicator - she effectively communicated her thoughts throughout the interview. Instead, it seems that there is a disconnect of expectations and norms between Moonbeam and her peers.

In addition to the disconnect between peers, the fear of being misinterpreted by instructors is detrimental to participants. Ray states, "\textit{I don't have a good metric for what other people think versus the things that I think,}" indicating that he navigates his problem-solving cautiously for fear of being misinterpreted, or that he will misinterpret a problem. He states that the constant fear of miscommunication means that he is "\textit{very paranoid that like I'm going to put an answer down that I actually understand that's going to be mistakenly classed as I cheated.}" This fear means that Ray navigates problem-solving very cautiously and with a belief that he cannot be successful at physics.

\begin{center}
    \textit{2. Interpreting Problem Statements}
\end{center}

We also found that the DEP arises in interpretation of problem statements, especially the intent behind problems. While solving a problem, Ray tells us that:
\begingroup
\addtolength\leftmargini{-0.16in}
\begin{quote}
"\textit{[When solving problems] you're supposed to look for the clues and then put them together, but I'm really, really good at misinterpreting vague things, so sometimes I misunderstand what the problem's asking and then end up doing an entirely different thing.}"
\end{quote}
\endgroup
This feeling of misinterpretation was brought up multiple times throughout all 3 participants' interviews. When probed about it, Ray indicates that the consistent feeling that he will misinterpret a problem leads him to believe that he is incapable of doing physics. In response to a question asking when he felt positive about being a physicist, Ray answers, "\textit{I genuinely cannot think of one instance.}"

\begin{center}
    \textbf{V. DISCUSSION AND CONCLUSION}
\end{center}

\begin{center}
    \textbf{A. Putting it All Together: Systemic Effects}
\end{center}

When asked if he had anything additional that he wanted people to know about being a neurodivergent physicist, Milo states,
\begingroup
\addtolength\leftmargini{-0.16in}
\begin{quote}
    "\textit{As much as you can fondly stereotype the physics bro as autistic or neurodivergent, you can also not so fondly stereotype the physics bro because the culture can certainly be toxic, or ... insensitive.}"
\end{quote}
\endgroup
When solving physics problems and discussing their experiences in physics, all 3 participants showed self-doubt, feelings of othering, and sometimes abject fear of being misinterpreted. When combined with previous analysis of neurodivergent student experiences that shows that physics culture is neuro-normative \cite{McDermott2024b}, the findings in this paper suggest that there may be a systemic effect of the DEP in physics problem-solving. 

That the 3 participants experience a negative impact from cross-neurotype communication breakdown (and that the breakdown could be systemic) is concerning, and this issue may be experienced by other neurodivergent students like the participants in this paper. In the following subsections, we present potential mitigation strategies for instructors to better support neurodivergent physics students.\\





\begin{center}
    \textbf{B. Breaking the cycle of DEP to expand student success}
\end{center}

As physics educators cognizant of the DEP and its effects, it is up to us to mitigate the harms that can arise from the communication breakdown. Indicative of the harms that can be done by a systemic DEP, Ray recognizes, "\textit{people tell me that I can [succeed in physics], but that doesn't exist.}" Ray does not believe that instructors are telling the truth when they offer supportive words or tell students they support multiple ways of doing physics problems. Counteracting this, Ray discusses a poster outside his introductory physics class that provides examples of non-white, non-male, and/or disabled physicists, saying,
\begingroup
\addtolength\leftmargini{-0.16in}
\begin{quote}
    "\textit{The thing that feels empowering to me is just seeing someone else doing it ... So even having one person who like told themselves "I can do it" and then succeeded is really, really nice to know. That, like, there are other people who are doing this differently than what's expected and the fact that they were able to do it means that it was a possibility that I can solve a problem.}" 
\end{quote}
\endgroup
Communicating that others like himself (scientists who do science differently than the norm) can succeed in their scientific aspirations gives Ray the self-efficacy that he can successfully solve a problem. Through providing evidence of non-white, non-male, disabled physicists' success, we can help students like Ray envision that success for themselves.


Students who do not fit within cisgender, abled, white male norms can be pushed to the margins in physics \cite{McDermott2024b,Santana2024,Ong2005,Zhang2025,Maries2025}. Oftentimes, succeeding in physics means passing the course and having working knowledge of the content. Having students achieve these metrics of success is important, and we also need to be mindful of \textit{which} and \textit{how} students are succeeding. Some of the student-centered metrics of success are explored in the literature through identifying positive shifts in physics identity, belonging, and collaboration skills \cite{Lock2015,Hazari2020}. All of these heavily rely on the mutual existence of empathy and effective communication. Encapsulating these ideas, Ray states that,
\begingroup
\addtolength\leftmargini{-0.16in}
\begin{quote}
    "\textit{One of the big moments for me, um, was I got like a homework back and the feedback was like `uh, you did all the math correct but all your numbers are incredibly off.' Like, I just didn't even care in that moment that all the numbers were off to have someone be like, `I acknowledge that you have done this thing correctly that you worked on and you were right,' like `you worked on it and you did it correctly.' I'm proud of that part. \textbf{I kind of completely changed the way that I looked at doing physics successfully}, of like being told you understand the concept was really, really nice}."
\end{quote}
\endgroup
This shift in communication is provided as an exemplar to show that we can work toward addressing the DEP in physics problem solving. Through recognizing diverse communication pathways and redesigning our pedagogy with diverse learners/communicators in mind, we have the power to effect lasting change in student's perceptions of success in physics.

\begin{center}
     \textbf{C. Implications for Physics Classrooms}
 \end{center}
 
The notion of "there is more than one way of solving a physics problem" is a common sentiment in physics instruction. However, it is clear from the data that this sentiment was not communicated to these 3 neurodivergent students (or not believed by, \textit{$\grave{a}$ la} Ray). To ground this research toward making change for physics learning spaces, we propose the following 8 tangible actions that can shift physics culture toward being more neuro-inclusive:
\begin{enumerate}
    \item Offer multiple means for feedback about level of understanding (e.g., verbally in class, anonymous survey, clicker questions, after class discussions),
    \item Model how to "search for clues" in problem statements that indicate a problem solving path,
    \item Avoid skipping steps when modeling problem solving in class,
    \item Model question asking \& follow-up questions,
    \item Provide rubrics for problem solving that allow for multiple pathways to solution,
    \item Validate different ways of solving problems (e.g., "I didn't think of it that way, and that's really cool!"),
    \item Verbalize expectations that students may not understand material right away (and provide examples from personal experience going through higher education),
    \item Provide examples of scientists who do science (or pursued their career in science) in non-normative ways.
\end{enumerate}

These suggestions serve to uncover aspects of the hidden curriculum that may not be immediately evident to neurodivergent students and provide a way for instructors to open a space for diverse forms of communication and thinking that neurodivergent students bring to the classroom. Our hope is that, through these suggestions, instructors can begin growing and develop instructional supports alongside neurodivergent learners to support all learners in the classroom.

\begin{center}
    \textbf{ACKNOWLEDGEMENTS}
\end{center}
This work was funded by the National Science Foundation (STEM Ed IPRF \#2411711) and the American Physical Society Forum on Education (FEd) Project Mini-Grant.

\newpage
\bibliographystyle{apsrev4-2}
\bibliography{bib}



\end{document}